\documentclass[aps,prb,amssymb,showpacs,superscriptaddress,preprint]{revtex4}
\usepackage{bm}
\usepackage{amsmath}
\usepackage{graphicx}
\DeclareMathOperator{\sgn}{sgn}
\DeclareMathOperator{\Ren}{Re}
\DeclareMathOperator{\Imn}{Im}
\begin{document}
\title{Electrical conductivity in graphene with point defects}
\author{Yuriy~V.~Skrypnyk}
\affiliation{G. V. Kurdyumov Institute of Metal Physics, 
             National Academy of Sciences of Ukraine
             Vernadsky Ave. 36,
             Kyiv 03680, Ukraine}
\author{Vadim~M.~Loktev}
\affiliation{Bogolyubov Institute for Theoretical Physics,
             National Academy of Sciences of Ukraine
             Metrolohichna Str. 14-b,
             Kyiv 03680, Ukraine}

\begin{abstract}
The electrical conductivity of graphene containing point defects is
studied within the binary alloy model in its dependence on the Fermi
level position at the zero temperature. It is found that the minimal
conductivity value does not have a universal character and corresponds
to the impurity resonance energy rather than to the Dirac point
position in the spectrum. The substantial asymmetry of the resulting
dependence of the conductivity on the gate voltage magnitude is
attributed as well to the very shift of the conductivity minimum to
the resonance state energy.   
\end{abstract}


\maketitle 

\section{Introduction} 
Graphene, a thermodynamically stable graphite monolayer, which has
been mechanically exfoliated for the first time only a few years
ago,\cite{gr1,gr2,gr3} is gaining considerable scientific
attention. This new material looks promising enough for a number of
important practical applications, some of which have been long hoped
for. While experimenters are targeted at engineering graphene based
devices in the not so distant future, graphene attracts theoreticians
as the first existing in the free state physical system, which can be
named two--dimensional (2D) without any reservations. Undoubtedly, so
far unique electronic properties of graphene were the most challenging
issue. These properties directly come up from the honeycomb lattice
with its two--atomic structure, inherent in a single atomic layer of
graphite. The lattice structure leads to the Dirac dispersion of
charge carriers, which makes up the core of studies devoted to
graphene.  

Transport properties of this material are, sure enough, of primary
importance for graphene--based electronics. In real crystals, transport
properties essentially depend on non--ideality of the system and on
interaction of carriers with other excitations. Below we are going to
focus on imperfections of graphene, and, in particular, on point
defects in it. This allows to rise a question on the spectrum of
delocalized carriers, on its dependence on the amount of defects, and,
eventually, on such a remarkable quantity as the minimal value of the
conductivity. 

Initially, two main features of the graphene conductivity were singled
out. First of all, the conductivity of graphene devices never dropped
below a certain value. Since this minimal value seemed not to vary
between different experimental samples, the origin of the universal
behavior of the minimal conductivity value has been extensively
searched for. These efforts shaped the famous \textit{minimum
conductivity} puzzle. The linear dependence of the conductivity on the
gate voltage made up the second feature. However, the minimal
conductivity value has been found soon to be strongly sample
dependent,\cite{nonun} and the effect of minimum conductivity has been
attributed to graphene's imperfections. In view of mentioned features
of the graphene conductivity, a qualitative difference between charged
impurities and point defects has been established.\cite{nomura} While
it has been demonstrated that charged impurities are able to yield the
required linear dependence of conductivity on the gate voltage, point
defects were shown to produce the sub--linear conductivity behavior,
and, consequently, ruled out as the conductivity limiting factor. The
concept of charged impurities as a main source of the scattering of
charge carriers in graphene has been thoroughly developed,
\cite{sarma,galits,adam} and convincingly compared to the experimental
data on graphene with deposited potassium atoms on its
surface.\cite{satur,fuhrer} At that, just a constant contribution to
the conductivity were ascribed to point defects when fitting the
experimental data. 

Even though the concept of charged impurities looks sounding,
experiments on graphene in ethanol environment seriously question the
dominant role of the Coulomb scatterers.\cite{nonc} In addition, the
conductivity asymmetry evident in measurements of graphene with
deposited potassium atoms has not received the proper explanation
yet.\cite{satur, fuhrer} While a moderate asymmetry can be attributed
to the disbalance of positively and negatively charged
impurities,\cite{nov} the marked asymmetry of conductivity dependence
on the gate voltage in graphene doped by transition metals manifests
the response that is different from the one, which is expected from
the charged impurity centers.\cite{kpi} The clearly sub--linear
character of conductivity curves corresponding to graphene samples
heavily doped by transition metals or lightly doped by the potassium
atoms only strengthen the overall impression that we are dealing with
the interplay of different types of disorder, and that each one of
them should receive a comprehensive treatment in an effort to grasp
the conductivity properties in graphene. As an example, an interplay
between charged impurities and point defects, which involves different
reaction of these two impurity types to screening effects, can be
employed to solve the dilemma of slightly varying minimal conductivity
value for graphene placed into a dielectric environment.\cite{fhr}

It must be stressed that in the mentioned experiments targeted at
finding the main scattering channel for carriers in graphene,
different adatoms were deposited on the surface of samples, which
otherwise were considered as pristine. In fact, here we are dealing
with two different issues again. One of them is determining what
actually limits the conductivity of graphene samples obtained by a
certain technique, and another one is analyzing the effect of
intentionally added impurities on the conductivity features. The
latter is closely related to the current tendency to functionalize
graphene by substitutionals, adatoms, or chemically active
groups. Regarding the conducting properties, such an adjustment of
graphene can proceed up to a possibility of the metal--insulator
transition, which has been successfully observed in graphene doped by
hydrogen atoms recently.\cite{eli} Similarly, graphene demonstrated
insulating behavior after irradiation by Ne ions, which is expected to
produce short--range defects.\cite{ion}

In view of that, before any complex models of impurity centers are
constructed, the simple ones should be properly examined in their
characteristic aspects. And the basic model of the point defect is
definitely among them, since it allows for a Mott transition in impure
graphene.\cite{skrloc}   

Below we are returning to the common in semiconductor physics model of
a binary alloy intending to examine what features of the graphene
conductivity it is capable to reproduce. Such short--range impurities
violate the electron--hole symmetry of the system and are naturally
providing for the conductivity asymmetry. This impurity model has been
studied either in a weak scattering limit, or in the unitary limit. We
are going to show that this model does exhibit some worthwhile features
in--between these two extremes. In particular, we demonstrate that
the effective conductivity minimum, in contrast to former studies,
corresponds not to the Dirac point of the spectrum, but to the energy
of a single impurity resonance, where the impurity scattering is the
strongest.     

\section{Model disordered system}
The host Hamiltonian for electrons in graphene is taken in the tight
binding approximation with hopping restricted to the nearest
neighbors, 
\begin{equation}
\mathbf{H}_{0}=t\sum\limits_{<\mathbf{n}\alpha,\mathbf{m}\beta>}%
c_{\mathbf{n}\alpha}^{\dagger}c^{\phantom{\dagger}}_{\mathbf{m}\beta}, 
\label{ham0}
\end{equation}
where $t\approx 2.7 \rm{eV}$ is the hopping parameter for the nearest
neighbors,\cite{t} $\bm{n}$ and $\bm{m}$ run over lattice cells,
$\alpha$ and $\beta$ enumerate two sublattices of the honeycomb atomic
arrangement, $c^{\dag}_{\bm{n}\alpha}$ and $c^{}_{\bm{n}\alpha}$ are
the electron creation and annihilation operators at the respective
lattice cites. Substitutional impurities are supposed to be
distributed evenly and in uncorrelated manner on lattice
sites. Presence of an impurity at a given lattice site is assumed to
be manifested only through a change in the respective on--site
potential of the tight--binding Hamiltonian. This type of impurity
perturbation fully corresponds to the conventional model of a binary
alloy with a diagonal disorder, which had been extensively used in
physics of real crystals, and sometimes is referred to as the Lifshitz
(or isotopic) model for historical reasons. The corresponding
Hamiltonian for a disordered graphene has the form,   
\begin{equation}
\bm{H} = \bm{H}_0 +\bm{H}_{imp},\qquad%
\bm{H}_{imp}=V_{L}{\sum_{\bm{n},\alpha}}%
\eta^{}_{\bm{n}\alpha}c^{\dag}_{\bm{n}\alpha}c^{}_{\bm{n}\alpha},
\label{fullH}
\end{equation}
where $V_{L}$ is the deviation of the potential at the impurity site,
and variable $\eta^{}_{\bm{n}\alpha}$ is unity with the probability
$c$ or zero with the probability $(1-c)$, which specifies $c$ as the
impurity concentration. The ``per carbon atom'' concentration $c$ can
be easily converted to the impurity coverage,
\begin{equation}
n_{im}=n_{0}c,
\end{equation}
where 
\begin{equation}
n_{0}=\frac{4}{\sqrt{3}a^{2}} \label{n0}
\end{equation}
is the inverse area per one carbon atom, and $a=0.246$ nm is the
lattice constant of graphene.

The Green's function of a disordered system,
\begin{equation}
\bm{\mathcal{G}}=(E-\bm{H})^{-1},
\end{equation}
after averaging over different impurity distributions,  
$\bm{G}=\biglb<\bm{\mathcal{G}}\bigrb>$, regains the translational
invariance and can be expressed by means of the Dyson equation,
\begin{equation}
\bm{G}=\bm{g}+\bm{g}\bm{\Sigma}\bm{G},
\end{equation}
where $\bm{\Sigma}$ is the self--energy, and
\begin{equation}
\bm{g}(E)=(E-\bm{H}_{0})^{-1}
\end{equation}
is the host Green's function. When the amount of introduced impurities
is moderate, it is possible to implement the modified propagator
method.\cite{langer} Within this approach, the self--energy is site
diagonal and identical on both sublattices,    
\begin{equation}
\bm{\Sigma}\approx\Sigma(E)\bm{I},\qquad%
\Sigma(E)\equiv\Sigma_{\bm{n}\alpha\bm{n}\alpha}(E)=\frac{cV_{L}}%
{1-V_{L}g_{\mathbf{n}\alpha\mathbf{n}\alpha}[E-\Sigma(E)]}, 
\label{mp}
\end{equation}
where $\bm{I}$ is the identity matrix. At a small impurity
concentration, $c\ll 1$, multiple occupancy corrections are not
significant. Thus, the method of modified propagator yields results
that are practically indistinguishable from the ones produced by the
conventional coherent potential approximation. 

The calculation of the diagonal element of the host Green's function
in the site representation $g_{\mathbf{n}\alpha\mathbf{n}\alpha}$,
which will be required for the subsequent analytical treatment of the
impurity problem, is given in Appendix. From here and on we are
choosing the bandwidth parameter $W$, see Eq.~(\ref{bw}), as the
energy unit. For the sake of clarity, we will use for the
dimensionless energy and the impurity potential following
designations:
\begin{equation}
\epsilon=\frac{E}{W},\qquad v=\frac{V_L}{W},
\end{equation}
Since only energies small compared to the bandwidth, i.e. those, at
which the linear dispersion holds in the host system, are considered,
the corresponding approximation to the dimensionless diagonal element
of the Green's function can be written as follows, 
\begin{equation}
W g_{\mathbf{n}\alpha\mathbf{n}\alpha}(E)\equiv g_{0}(\epsilon)%
\approx 2\epsilon\ln\left|\epsilon\right|-i\pi\left|\epsilon\right|,%
\quad\left|\epsilon\right|\ll 1. \label{gf}
\end{equation}

It is not difficult to see that the diagonal element of the Green's
function (\ref{gf}) looks similar to the properly scaled diagonal
element obtained within the model of massless Dirac fermions for the
electron spectrum in graphene. Thus, despite the fact, that the
intervalley scattering is, for sure, taken into account in
Eq.~(\ref{mp}) explicitly, the net result for the single--site
scattering will not be qualitatively different from the one for the
frequently used model of massless Dirac fermions, in which only one
Dirac cone is retained. This resemblance follows from the single--site
character of the impurity perturbation (see Eq.~(\ref{fullH})). At
that, the validity of the modified propagator method (\ref{mp}) is
limited by the scatterings on impurity clusters,\cite{skrloc,persh}
which contribution to the self--energy will be monitored in what
follows.
  
\section{Renormalized energy phase and conductivity}

In order to make the self--consistency condition (\ref{mp}) more
tractable, a regular substitution can be made,
\begin{equation}
\epsilon-\Sigma(\epsilon)=\varkappa \exp(i\varphi),\quad %
\varkappa>0,\quad 0<\varphi<\pi, \label{subs}
\end{equation}
which singles out the phase of the renormalized energy
$\epsilon-\Sigma(\epsilon)$. This phase diminishes from $\pi/2$ to
zero inside the conduction band and rise from $\pi/2$ to $\pi$ within
the valence band when moving away from the Dirac point position. With
the help of the obtained above expression (\ref{gf}) for the diagonal
element of the Green's function and the substitution (\ref{subs}), the
imaginary part of Eq.~(\ref{mp}) can be reduced as follows, 
\begin{multline}
cv^{2}\left[2\ln\varkappa+(2\varphi-\pi)\cot\varphi\right]+%
\left[1-v\varkappa(2\ln\varkappa\cos\varphi-(2\varphi-\pi)%
\sin\varphi\right)]^{2}+ \\
+\left[v\varkappa(2\ln\varkappa\sin\varphi+%
(2\varphi-\pi)\cos\varphi\right)]^{2}=0. \label{sim} 
\end{multline}
Provided that the impurity perturbation strength $v$ and the impurity
concentration $c$ are fixed, this equation establishes a
correspondence between the renormalized energy modulus $\varkappa$ and
its phase $\varphi$. For those $\varkappa$ that are exceeding a
certain threshold magnitude, which is, indeed, determined by the
impurity concentration and the perturbation strength, this equation
always has two different solutions with respect to the phase
$\varphi$. One of them ($\varphi<\pi/2$) belongs to the conduction
band, while the other ($\varphi>\pi/2$) lies within the valence
band. The literal carrier energy that corresponds to a given
renormalized energy is determined by the real part of Eq.~(\ref{mp}),
\begin{multline}
\epsilon=\varkappa\cos\varphi+ \\
+\frac{cv\left[1-v\varkappa(2\ln\varkappa\cos\varphi-%
(2\varphi-\pi)\sin\varphi)\right]}%
{\left[1-v\varkappa(2\ln\varkappa\cos\varphi-%
(2\varphi-\pi)\sin\varphi)\right]^{2}+%
\left[v\varkappa(2\ln\varkappa\sin\varphi+%
(2\varphi-\pi)\cos\varphi)\right]^{2}}. \label{sim2} 
\end{multline}
Taken together, the last two equations, (\ref{sim}) and (\ref{sim2}),
are making up a set, which implicitly specifies the dependence of the
renormalized energy phase $\varphi$ on the carrier energy $\epsilon$.

Since the procedure required to calculate the self--energy is already
outlined, it is possible to employ the Kubo expression for the
conductivity of a disordered graphene at the zero
temperature,\cite{hirash} 
\begin{equation}
\tilde{\sigma}_{cond}=\frac{4e^{2}}{\pi h}%
\left\{1+\Biggl[\frac{\epsilon_{F}-\Ren\Sigma(\epsilon_{F})}%
{-\Imn\Sigma(\epsilon_{F})}%
+\frac{-\Imn\Sigma(\epsilon_{F})}%
{\epsilon_{F}-\Ren\Sigma(\epsilon_{F})}\Biggr]%
\arctan\Biggl[\frac{\epsilon_{F}-\Ren\Sigma(\epsilon_{F})}%
{-\Imn\Sigma(\epsilon_{F})}\Biggr]\right\}, 
\end{equation}
where $\epsilon_{F}$ is the Fermi energy. By means of the substitution
(\ref{subs}), which was used above to simplify the self--consistency
condition for the self--energy (\ref{mp}), the above expression can be
significantly reduced,    
\begin{equation}
\tilde{\sigma}_{cond}%
=\left(\frac{e^{2}}{h}\right)\sigma_{cond}, \qquad
\sigma_{cond}=\frac{2}{\pi}\left[1+%
(\cot\varphi_{F}+\tan\varphi_{F})\Bigl(\frac{\pi}{2}%
-\varphi_{F}\Bigr)\right], \label{kub}
\end{equation}
where $\varphi_{F}$ is the renormalized energy phase at the Fermi
level, and the dimensionless conductivity $\sigma_{cond}$, which will
be used onwards, is singled out. It should be emphasized that the
dimensionless conductivity $\sigma_{cond}$ depends on the renormalized
energy phase $\varphi$ alone. In the same way, the well--known
Ioffe--Regel criterion,\cite{Ioffe} which is commonly used to separate
extended states in a disordered system, and the applicability
criterion of the modified propagator method can both be expressed
through the same renormalized energy phase.\cite{skr} It has been
shown that with varying the renormalized energy phase the modified
propagator approximation validity violation and the indication of the
state localization by the Ioffe-Regel criterion are occurring
simultaneously for those states, which energies fall inside the host
band of a disordered system.\cite{skr,skrloc} Certainly, it is not
conceptually correct to expect that the Ioffe--Regel criterion can be
used to pinpoint precisely the mobility edge position in a disordered
system. Similarly, there should be no sharp boundaries between
those spectral regions, in which the modified propagator method is
applicable, and those ones, in which it is not. Nevertheless, there
are strong arguments supporting the estimation that the mobility edge
in a disordered system should be located at those energy, at which the
renormalized energy phase is close to $\pi/6$ for the conduction band,
and, respectively, to $5\pi/6$ for the valence band. Thus, in those
spectral intervals, inside which states are anticipated to be localized
according to the Ioffe--Regel criterion, neither the Kubo formula
(\ref{kub}) has any relevance, nor the modified propagator method is
reliable. On the contrary, the approach outlined above is consistent
in the spectral domains occupied with extended states, where the
renormalized energy phase $\varphi$ is either small (for the
conduction band) or close to $\pi$ (for the valence one).

\section{Conductivity in different scattering regimes}

\subsection{Weak scatterers}

When the impurity perturbation strength is moderate ($|v|<1$), it is
possible to take an advantage of the renormalized energy phase
smallness (or its closeness to $\pi$) and construct a correspondent
approximate solution of Eq.~(\ref{sim}),
\begin{equation}
\theta\approx\frac{\pi c v^{2}}%
{(1\mp 2v\varkappa\ln\varkappa)^{2}+(\pi v\varkappa)^{2}%
+2cv^{2}(1+\ln\varkappa)}, \qquad \theta\ll 1, \label{ap1}
\end{equation}
where $\theta$ stands for $\varphi$ inside the conduction band, and
for $\pi-\varphi$ inside the valence band. The sign in the denominator
also switches from a minus to a plus when moving from the conduction
band to the valence band. Obviously, the renormalized energy phase is
close to $\pi/2$ in a narrow interval of energies around the shifted
Dirac point, and thus the above approximation is not valid inside this
region. However, the transition of the renormalized energy phase
from small values to values that are close to $\pi$ is very fast. This
transitional region is, in fact, exponentially narrow and, for certain
reasons, should be treated separately, as it will be explained in
detail below. 

It is not difficult to check that in this scattering regime ($|v|<1$)
the effective shift of states along the energy axis, which is given by
the real part of the self--energy $\Ren\Sigma(\epsilon)$, is nearly
constant in the whole domain under consideration
($|\epsilon|\ll1$). Therefore, as a first approximation, one can take
\begin{equation}
\pm\varkappa\approx\epsilon-cv, \label{ap2}
\end{equation}
where the sign is varying according to the current band as above, so
that $\varkappa$ always remains positive, as it should do. The
expression for the conductivity, Eq.~(\ref{kub}), can be also
simplified utilizing the smallness (or closeness to $\pi$) of the
renormalized energy phase, 
\begin{equation}
\sigma_{cond}\approx\frac{1}{\theta}, \qquad \theta\ll 1, \label{ap3}
\end{equation}   
where linear terms and terms of the higher order in $\theta$ are
omitted. All these approximations, Eqs.~(\ref{ap1})--(\ref{ap3}), can
be combined into the final expression for the dimensionless
conductivity,
\begin{equation}
\sigma_{cond}\approx\frac{[1-2v(\epsilon_{F}-cv)%
\ln|\epsilon_{F}-cv|]^{2}+[\pi v(\epsilon_{F}-cv)]^{2}}%
{\pi cv^{2}}+\frac{2}{\pi}[1+\ln|\epsilon_{F}-cv|], \label{vsm}
\end{equation}
which fits well the conductivity, calculated numerically by
Eqs.~(\ref{sim}),(\ref{sim2}), and (\ref{kub}), throughout the whole
considered interval of energies ($|\epsilon|\ll 1$).

\begin{figure}
\includegraphics[width=0.96\textwidth]{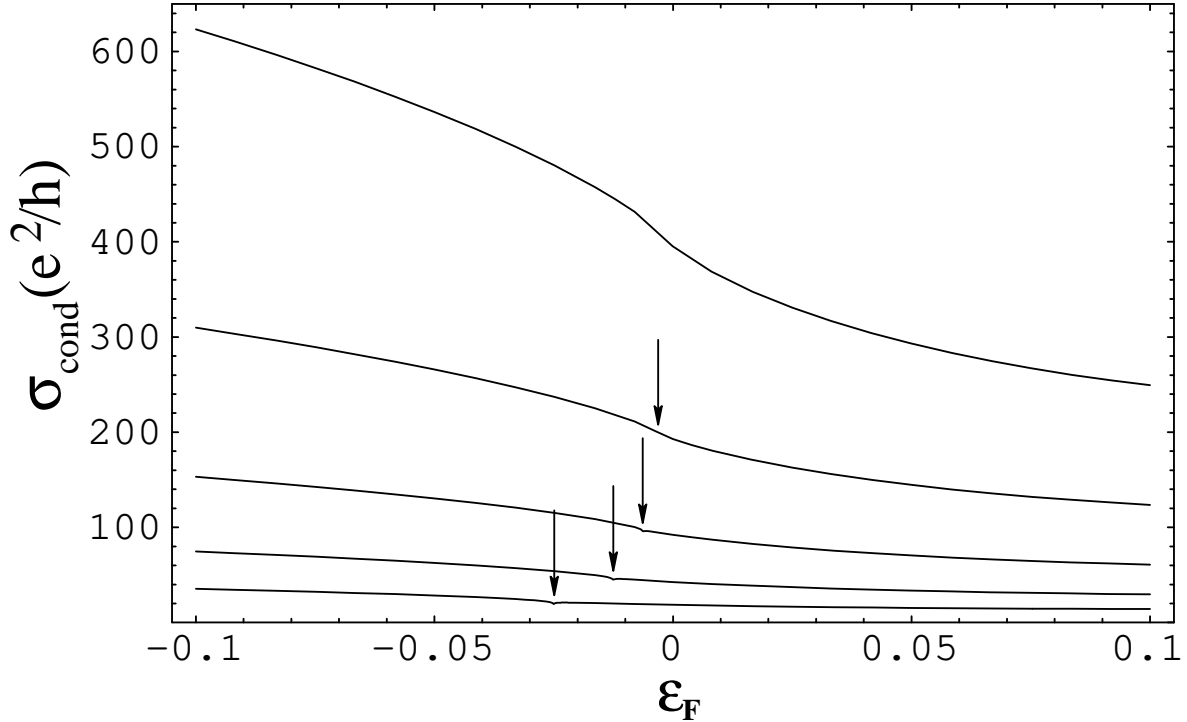}
\caption{\label{f1} Conductivity of graphene with point defects 
\textit{vs} Fermi energy for $v=-0.5$ and concentrations 
$c=0.1/2^{n}$, $n=1 \dots 5$. Arrows point at positions of narrow
dips.} 
\end{figure}

As follows from (\ref{vsm}), the conductivity is gradually diminishing
with increasing the Fermi energy $\epsilon_F$ from the valence band to
the conduction band for a negative impurity perturbation ($v<0$) and
vice versa. The conductivity of graphene calculated by
Eqs.~(\ref{sim}), (\ref{sim2}), and (\ref{kub}) without any additional
approximations at different concentrations of point defects is plotted
against the Fermi energy in Fig.~\ref{f1} for the case of moderate
impurity perturbation ($|v|<1$). On the whole, the dependence of the
conductivity on the Fermi energy is smooth and almost featureless,
while being strongly asymmetric against the shifted Dirac point
position. The only exception from the monotonic behavior of the
conductivity can be observed in a close vicinity of the Dirac point,
where the curve manifests a sharp dip, which is barely discernible in
the curves corresponding to high impurity concentrations. If to trust
the results yielded by the modified propagator method all the way down
to the Dirac point, the conductivity at its very tip should drop to
$4/\pi$. This directly corresponds to the notorious theoretical
magnitude of the universal minimum conductivity in the inhomogeneous
graphene, which has been widely debated within the so--called ``missed
$\pi$'' discourse. However, the modified propagator method is not
applicable in the Dirac point neighborhood. It can be
shown\cite{skrloc,skrloc2} that this approximation is not reliable in
the interval 
\begin{equation}
|\epsilon-cv|\lesssim \exp(-\frac{1}{4cv^{2}}-1).
\end{equation} 
Therefore, as it was outlined in the previous section, the Kubo
formula is also ineffective in this interval, and the obtained
conductivity magnitude at the Dirac point has no physical
meaning. Still, the width of the energy interval, in which the
analytical approach fails, is exponentially small compared to the
bandwidth. Since the corresponding dip on the conductivity curve is so
narrow, it should be averaged out at realistic sample temperature, or
by means of any other broadening mechanism.  

Consequently, the presence of the sharp dip on the conductivity curve
can be neglected, and, probably, should never come out in actual
experiments. The asymmetry of the conductivity dependence on the Fermi
energy arises from the presence of the severely smeared out impurity
resonance, which enhances the impurity scattering. Indeed, the smooth
character of the conductivity curve does not resemble the
experimentally observed check mark shape. However, if this check mark
shape is caused by another dominating type of impurities, weakly
scattering point defects undoubtedly can contribute to the asymmetry
of the conductivity curve.   

\subsection{Strong scatterers}
   
In the limit of the strong impurity scattering, $|v|\gg1$, situation
is completely different. When the impurity potential is large
compared to the bandwidth, a well--defined resonance state is
manifested in the electron spectrum.\cite{skrloc} In the limit of a
strong impurity potential, the resonance state energy $\epsilon_{r}$
is determined by the Lifshitz equation,     
\begin{equation}
1\approx v\mathop{\mathrm{Re}}g_{0}(\epsilon_{r})%
\approx 2v\epsilon_{r}\ln\left|\epsilon_{r}\right|, \label{le}
\end{equation}
while the resonance state damping is given by
\begin{equation}
\Gamma_{r}\approx\frac{\pi|\epsilon_{r}|}{2|1+\ln|\epsilon_{r}||}.
\label{gam}
\end{equation}
For the resonance state to be well--defined, the condition
\begin{equation}
\gamma_{r}\equiv\frac{\Gamma_{r}}{|\epsilon_{r}|}\approx\frac{\pi}%
{2\left|1+\ln|\epsilon_{r}|\right|}\ll 1 \label{wd}
\end{equation} 
must be met. Thus, one should have $|\ln|\epsilon_{r}||\gg1$, which
corresponds to a strong impurity perturbation and a resonance energy
located close to the Dirac point.

The qualitative difference of the strong impurity perturbation case
resides not only on the presence of a resonance state in the
spectrum, but mainly on the fact that in this case the electron
spectrum undergoes a radical rearrangement. That is, with increasing
impurity concentration a quasigap filled with localized states opens
up around the resonance state energy.\cite{skrloc,skrloc2,persh} 
There exists a certain critical concentration of impurities,
\begin{equation}
c_{r}\sim-\frac{1}{2v^{2}\ln(\zeta/|v|)}, \qquad \zeta\sim 1,
\end{equation}   
which is determined by the mutual spatial overlap of individual
impurity states. When impurity concentration exceeds the critical
concentration $c_{r}$ of the spectrum rearrangement, the width of the
quasigap starts to increase rapidly with increasing impurity 
concentration as $\sqrt{-2c/\ln c}$.\cite{skrloc,skrloc2,persh}
Certainly, neither the modified propagator method nor the Kubo
expression for the conductivity will work inside this
quasigap. Therefore, we will consider only those impurity
concentrations that are less than the critical one ($c<c_{r}$) in the
case of the strong impurity potential. Vacancies are frequently
modeled by point defects with infinite impurity potentials
$v$. Because of this, the critical concentration $c_{r}$ for vacancies
in graphene is zero. In other words, the spectrum rearrangement is
already over for any concentration of vacancies. Therefore, vacancies
are out of the scope of the present study. 
 
\begin{figure}
\includegraphics[width=0.96\textwidth]{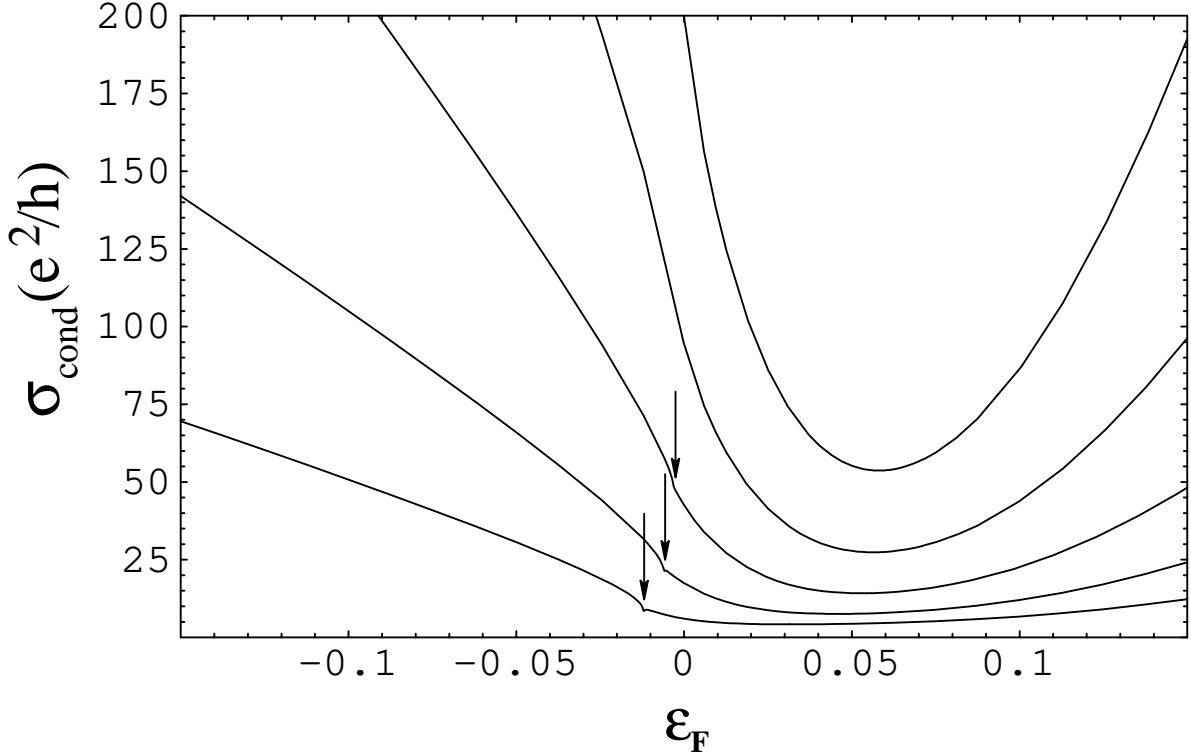}
\caption{\label{f2} Conductivity of graphene with point defects
\textit{vs} Fermi energy for $v=-2$ and concentrations
$c=c_{r}/2^{n}$, $n=1 \dots 5$, $c_{r}\approx 0.012$. Arrows point at
positions of narrow dips.} 
\end{figure}

\begin{figure}
\includegraphics[width=0.96\textwidth]{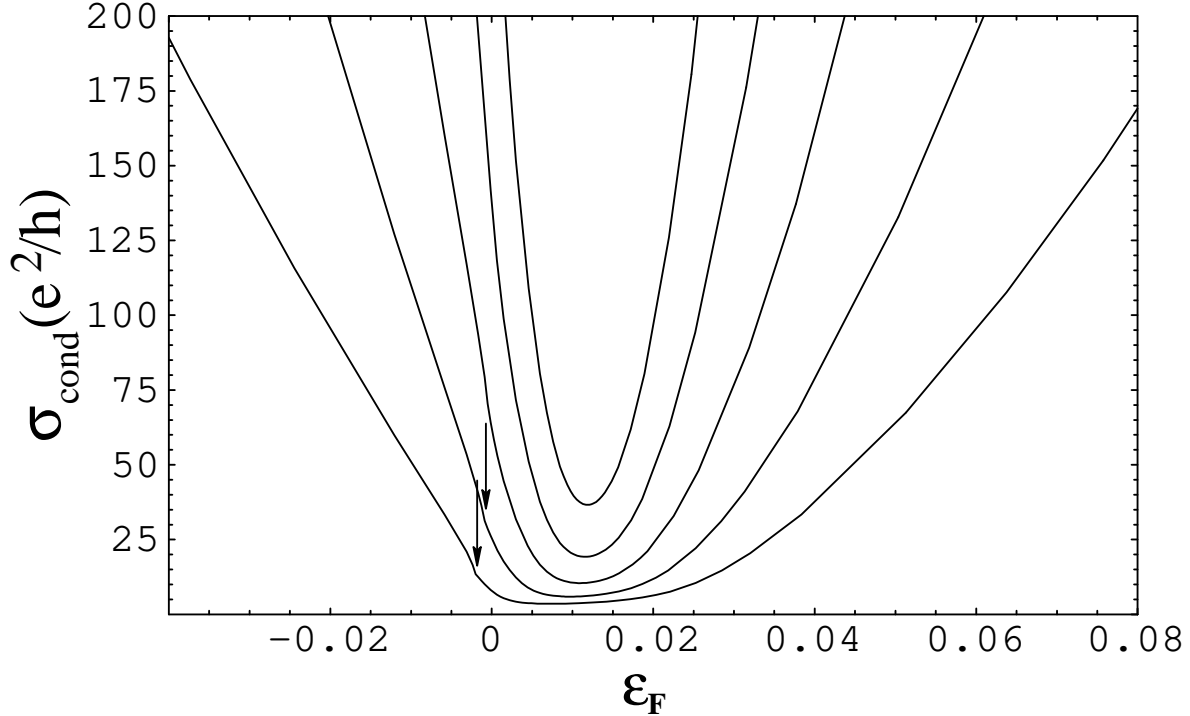}
\caption{\label{f3} Conductivity of graphene with point defects
\textit{vs} Fermi energy for $v=-8$ and concentrations
$c=c_{r}/2^{n}$, $n=1 \dots 5$, $c_{r}\approx 0.0005$. Arrows point at
positions of narrow dips.}
\end{figure}

The conductivity calculated directly by Eqs.~(\ref{sim}),
(\ref{sim2}), and (\ref{kub}) at different concentrations of point
defects is shown in Figs.~\ref{f2} and \ref{f3} for a not so
excessive ($v=-2$) and for a reasonably strong ($v=-8$) impurity
potential, respectively. The Dirac point shift, which occurs along
with the impurity concentration increase, is not so pronounced,
because $c_{r}v\sim 1/v$. Like in the case of the weak impurity
potential, there is a sharp dip in the conductivity curve located at
the Dirac point. The hint of this dip can be seen in the figures at
concentrations that are approaching the critical one. Nevertheless,
the presence of this dip should be neglected by the same arguments as
above.  
  
What really distinguishes the strong impurity perturbation case is the
presence of the clear minimum on the conductivity curve, which is
located at the energy of the impurity resonance state. This is
understandable, since the impurity scattering is the strongest around
the resonance energy. The width of this minimum corresponds to the
resonance state broadening, and, therefore, this minimum is not as
sharp as the minimum at the Dirac point. Overall, the conductivity
curve acquires a quasi--parabolic form, which is particularly
well--pronounced at lower impurity concentrations. In addition, the
conductivity curve appears more symmetric for a larger impurity
potential. 

With increasing the impurity concentration, the concentration
broadening of the resonance state also increases. At the impurity
concentrations that are close to the critical one, the broadening of
the resonance state is as wide as the distance from the resonance
energy to the Dirac point. This widening of the resonance broadening
area along with the tendency of states toward localization inside it
are manifested by the apparent flattening of the conductivity curve
around the resonance energy at $c\sim c_{r}$. Outside the domain of
the concentration broadening, the approximate expression for the
conductivity is even simpler than before, 
\begin{equation}
\sigma_{cond}\approx\frac{[1-2v\epsilon_{F}%
\ln|\epsilon_{F}|]^{2}+[\pi v\epsilon_{F}]^{2}}%
{\pi c v^{2}}. \label{vsm2}
\end{equation}
It is not difficult to see from Eq.~(\ref{vsm2}) that in the unitary
limit, 
\begin{equation}
\sigma_{cond}\rvert^{\phantom{A}}_{v\rightarrow\infty}\approx%
\frac{\epsilon_{F}^{2}[(2\ln|\epsilon_{F}|)^{2}%
+\pi^{2}]}{\pi c}, \label{uni}
\end{equation}
which corresponds to the known result for vacancies in
graphene.\cite{stb} As was stated above a quasigap around the Dirac
point should be present at any impurity concentration in the unitary
limit of the impurity perturbation. The approximation (\ref{uni}) is
valid only outside of this quasigap.

For the finite impurity perturbation, the expression (\ref{vsm2}) also
can not be used close to the resonance energy. In order to obtain the
minimum value of the conductivity, it is required to know the
magnitude of the renormalized energy phase at the resonance
energy. Its concentration dependence, $\varphi_{r}(c)$, follows from
the self--consistency condition (\ref{sim}). The second term in this
equation nullifies by the very definition of the resonance energy
Eq.~(\ref{le}). The remaining two terms constitute the relation: 
\begin{equation}
c=-2\epsilon_{r}^{2}(c)\tan\varphi_{r}(c)\Bigl[\ln\Bigl|%
\frac{\epsilon_{r}(c)}{\cos\varphi_{r}(c)}\Bigr|\tan\varphi_{r}(c)%
+\varphi_{r}(c)-\frac{\pi}{2}\Bigr]. \label{cself}
\end{equation} 
This expression can be significantly simplified by taking into account
that introduced earlier symmetric phase $\theta$ is small at low
impurity concentrations,
\begin{equation}
c\approx\pi\epsilon_{r}^{2}(c)\theta_{r}(c)%
\Bigl[1+\frac{\theta_{r}(c)}{\gamma_{r}}\Bigr],\qquad %
\theta_{r}(c)\ll 1. \label{cselfsim}
\end{equation}

Dependence of the resonance energy on the concentration
$\epsilon_{r}(c)$ is very weak for the strong impurity perturbation
and can be neglected. By setting the resonance energy in
Eq.~(\ref{cselfsim}) to its value for the isolated impurity, this
equation can be solved for the renormalized energy phase at the
resonance $\theta_{r}$, 
\begin{equation}
\theta_{r}(c)\approx\frac{\gamma_{r}}{2}%
\left(\sqrt{1+\frac{c}{c^{*}}}-1\right), \qquad %
c^{*}=\frac{\pi\epsilon_{r}^{2}\gamma_{r}}{4} \label{phase}
\end{equation}

\begin{figure}
\includegraphics[width=0.96\textwidth]{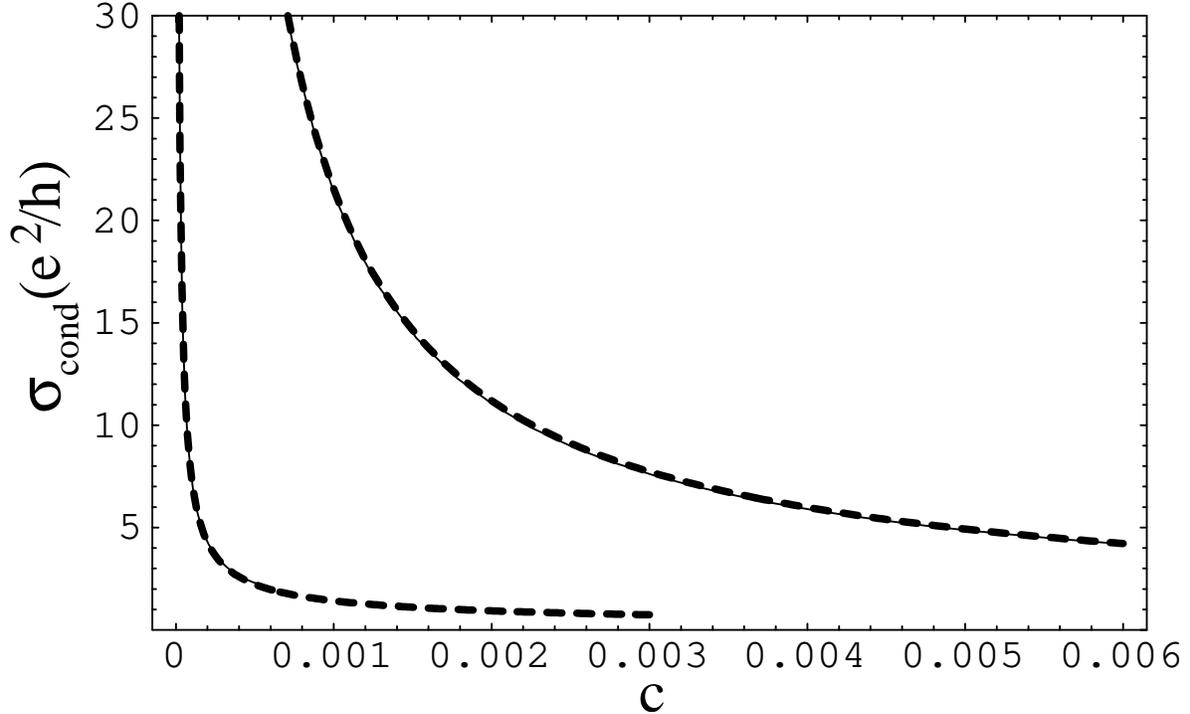}
\caption{\label{f4} The minimum conductivity \textit{vs} impurity
concentration for $v=-2$ (upper curve) and for $v=-8$ (lower
curve). Corresponding analytical approximations are given in dashed
lines.}
\end{figure}

The minimum conductivity at the resonance energy is then given by
Eq.~(\ref{ap3}). Indeed, it does not again have the universal
character and varies with impurity concentration. The minimum value of
conductivity calculated numerically with the help of Eqs.~(\ref{sim}),
(\ref{sim2}) and (\ref{kub}) is plotted against the impurity
concentration in Fig.~\ref{f4} for two different values of the
impurity potential. Initial fast minimum conductivity drop, which
occurs with increasing the impurity concentration, is followed by a
considerable flattening of the curve. The manifested saturation--type
behavior of the minimum conductivity concentration dynamics
qualitatively corresponds to the observed data.\cite{satur} According
to Eqs.~(\ref{phase}) and (\ref{ap3}), at small impurity
concentrations, $c\ll c^{*}$, the minimum conductivity of graphene
with point defects is proportional to $1/c$, which is similar to the
case of charged impurities.\cite{charg} However, at higher impurity
concentrations, $c\gg c^{*}$, the minimum conductivity for point
defects falls down more slowly, namely as $1/\sqrt{c}$. Thus, if both
expressions for the minimum conductivity are fitted to each other in
the low concentration limit, then the one corresponding to the point
defects should yield significantly larger values of the minimum
conductivity at $c\gg c^{*}$.

\section{Conductivity asymmetry}

\begin{figure}
\includegraphics[width=0.96\textwidth]{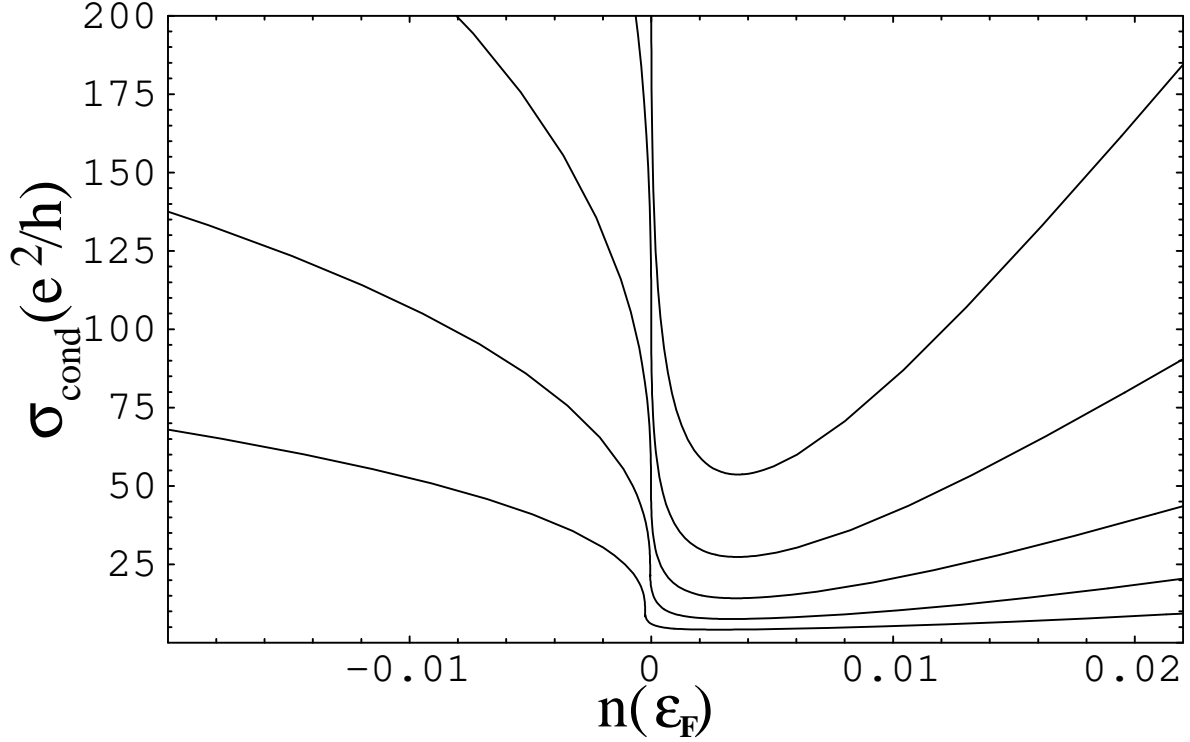}
\caption{\label{f6} Conductivity of graphene with point defects
\textit{vs} number of carriers for $v=-2$ and concentrations
$c=c_{0}/2^{n}$, $n=1 \dots 5$, $c_{0}\approx 0.012$.} 
\end{figure}

\begin{figure}
\includegraphics[width=0.96\textwidth]{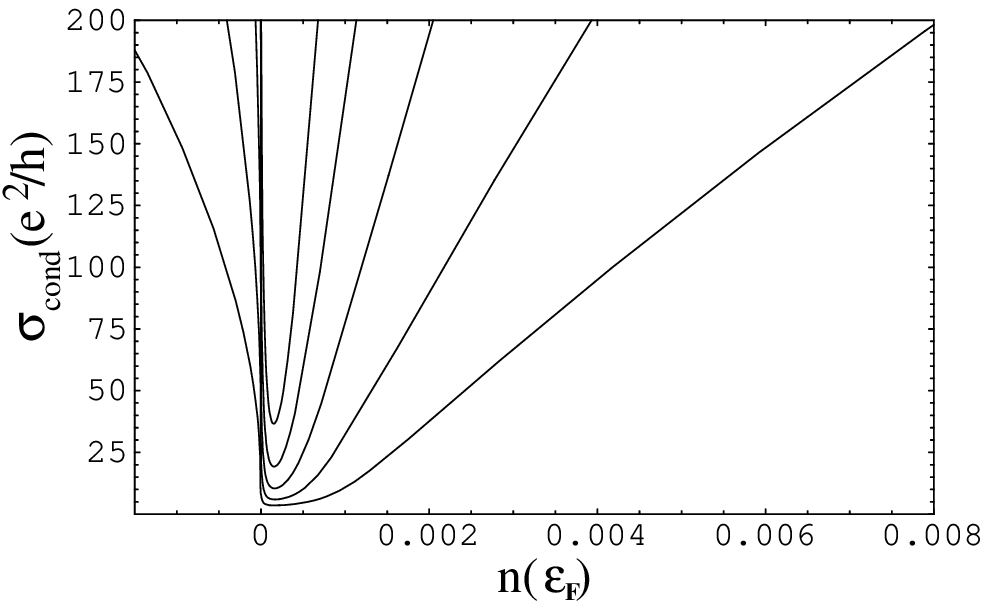}
\caption{\label{f7} Conductivity of graphene with point defects
\textit{vs} number of carriers for $v=-8$ and concentrations
$c=c_{0}/2^{n}$, $n=1 \dots 5$, $c_{0}\approx 0.0005$.} 
\end{figure}

The conductivity of graphene devices is usually measured against the
applied gate voltage. Since the gate voltage controls the carrier
density in the graphene sample, these experimental curves can be
simulated by plotting the conductivity as a function of the number of
occupied states. Leaving out the irrelevant constant, and taking into
account two actual sublattices, the number of occupied states can be
written as follows,   
\begin{equation}
n(\epsilon_{F})=-\frac{2}{\pi}\int_{0}^{\epsilon_{F}}\Imn(\{\epsilon-%
\Sigma(\epsilon)\}\{2\ln[\epsilon-\Sigma(\epsilon)]-i\pi\})d\epsilon.
\label{num}
\end{equation}
The conductivity of impure graphene, calculated as before by
Eqs.~(\ref{sim}), (\ref{sim2}) and (\ref{kub}), is plotted in
Figs.~\ref{f6} and \ref{f7} for both chosen strengths of impurity
perturbation against the number of occupied states, which is given by
Eq.~(\ref{num}). The number of occupied states is calculated by the
numerical integration. The change in the introduced magnitude
$\Delta n(\epsilon_{F})$ can be easily related to the respective
change in the carrier density $\Delta n$,
\begin{equation}
\Delta n=n_{0}\Delta n(\epsilon_{F}),
\end{equation}
where $n_{0}$ is defined by Eq.~(\ref{n0}). In the usual experimental
setup, the carrier density depends linearly on the gate voltage
$V^{g}$, so $\Delta n=\chi_{V} \Delta V^{g}$, where 
$\chi_{V}\approx 7.3\times 10^{10}\ \textrm{cm}^{-2}\textrm{V}^{-1}$.
Thus, when the gate voltage is varying in the window of 
$\pm 100\ \textrm{V}$, the value of $n(\epsilon_{F})$ varies by
$0.004$. 
 
It is visible from Figs.~\ref{f6} and \ref{f7} that the calculated
conductivity dependence on the gate voltage is highly
asymmetric. Similar asymmetric character of the conductivity curve has
been already reported elsewhere for the graphene with point
defects.\cite{stauber} While the expected slightly sublinear behavior
of the impure graphene conductivity is readily reproduced, the
asymmetry of the curve appears to be a bit on the extreme
side. Although such a strong asymmetry is sometimes reported for a
graphene with deposited adatoms,\cite{kpi} its origin for the point
defects is to be understood. In order to proceed in this direction,
the conductivity can be expanded into a series in the vicinity of its
minimum. Since it has been reasoned above that the conductivity
reaches its minimum value at the resonance energy $\epsilon_{r}$ for
the strong impurity perturbation, the expansion is straightforward,   
\begin{equation}
\sigma_{cond}\approx\sigma_{cond}^{0}+\chi_{r}(\epsilon_{F}-%
\epsilon_{r})^{2}, \qquad \chi_{r}>0 \label{st}
\end{equation}
where $\sigma_{cond}^{0}$ is the minimum value of the conductivity,
and $\chi_{r}$ is some constant.

In the current paper we restrict ourselves for the strong scattering
regime to those impurity concentrations that are less than the
critical concentration $c_{r}$ of the spectrum rearrangement. In
this case, the density of states is not considerably distorted by the
presence of defects. Because of the estimative character of this
arithmetic, it is quite sufficient to assume that the density of
states remains completely unchanged, i.e. identical to the host
system, except of the rigid shift of both bands to a new Dirac point
$\epsilon_{D}$,  
\begin{equation}
V^{g}_{\phantom{D}}\approx V^{g}_{D}+\chi_{D}(\epsilon_{F}-%
\epsilon_{D})^{2}\sgn(\epsilon_{F}-\epsilon_{D}), \qquad \chi_{D}>0, 
\label{vd}
\end{equation}     
where $V^{g}_{D}$ is those magnitude of the gate voltage
$V^{g}_{\phantom{D}}$, at which the Fermi level comes to the Dirac
point of the spectrum, and $\chi_{D}$ is some constant. This equation
can be easily solved for the Fermi energy, 
\begin{equation}
\epsilon_{F}\approx\epsilon_{D}+\sqrt{\frac{|V^{g}_{\phantom{D}}%
-V^{g}_{D}|}{\chi_{D}}}\sgn(V^{g}_{\phantom{D}}-V^{g}_{D}).
\end{equation}
Substituting this result to the expansion (\ref{st}), one can obtain: 
\begin{equation}
\sigma_{cond}\approx \sigma_{cond}^{0}+\frac{\chi_{r}}{\chi_{D}}%
\left[\sqrt{|V^{g}_{\phantom{D}}-V^{g}_{D}|}%
\sgn(V^{g}_{\phantom{D}}-V^{g}_{D})-\sqrt{\chi_{D}}%
(\epsilon_{r}-\epsilon_{D})\right]^{2}.
\end{equation}
It is evident from this relation, that the conductivity asymmetry
arise from the shift expressed by the second term in the square
brackets. If the conductivity reaches its minimum precisely at the
Dirac point, then the conductivity dependence on the Fermi energy is
linear and symmetric. However, we have demonstrated that the
conductivity minimum is attained at the impurity resonance energy,
which is essentially different from the Dirac point energy. This very
difference does form the ground for the substantial conductivity
asymmetry.  

\section{Conclusion}
It is demonstrated that there are two scattering regimes, which
characterize the behavior of the conductivity in graphene with point
defects. In the weak scattering regime, i.e. when the impurity
perturbation strength is less than the bandwidth, the dependence of
the conductivity on the Fermi energy is monotonic and asymmetric,
which can contribute to the observed conductivity asymmetry, when
point defects does not dominate other sources of scattering. However,
in the strong scattering regime, i.e. when the impurity potential
exceeds the bandwidth, the conductivity caused by point defects
manifests a distinctive minimum in its dependence on the Fermi
energy. This minimum, in contrast to a majority of anticipations,
corresponds not to the Dirac point of the spectrum, but to the
impurity resonance energy. In this regime, the asymmetry of the
conductivity dependence on the Fermi energy is noticeable. What is
more, the pronounced asymmetry of the corresponding dependence of the
conductivity on the gate voltage is caused by the very shift of the
conductivity minimum from the Dirac point to the impurity resonance
energy. Despite the basic nature of the considered impurity model, it
can qualitatively capture the essential features of the impure
graphene conductivity manifested in experiments. Thus, one can expect
that increasing the number of parameters characterizing the point
defect will permit to approach closer to the quantitative description
of conductivity features in graphene with point defects. 
 
\begin{acknowledgments}
Authors are grateful to V.~P.~Gusynin for valuable discussions. This
work was supported by the SCOPES grant $\mathcal{N}^{\circ}$~%
IZ73Z0-128026 of Swiss NSF, by the SIMTECH grant
$\mathcal{N}^{\circ}$~246937 of the European FP7 program, and by the
Program of Fundamental Research of the Department of Physics and
Astronomy of the National Academy of Sciences of Ukraine. 
\end{acknowledgments}

\appendix*
\section{Diagonal element of Green's function}
The straightforward expression for the diagonal element of the host
Green's function reads:
\begin{equation}
g_{\bm{n}\alpha\bm{n}\alpha}(E)=%
\frac{1}{S_{BZ}}\int\frac{E}{E^{2}-{E}^{2}(\bm{k})}d\bm{k},
\label{orig}
\end{equation} 
where the integration is carried over the entire Brillouin zone, which
has the area
\begin{equation}
S_{BZ}=\frac{8\pi^{2}}{\sqrt{3}a^{2}},
\end{equation}    
and ${E}(\bm{k})$ is the unperturbed dispersion relation corresponding
to the host Hamiltonian (\ref{ham0}).
 
In practical situations the Fermi level in graphene is located, nearly
unavoidably, in a narrow spectral region, in which the dispersion is
linear with a good accuracy. Near each of the two inequivalent Dirac
points, the dispersion relation $E(\bm{k})$ can be expanded,
\begin{equation}
E(\bm{k'})\approx\pm v_{F} k',
\end{equation}
where $\bm{k'}$ is taken relative to the corresponding Dirac point,
and   
\begin{equation}
v_{F}=\frac{\sqrt{3} a t}{2},
\end{equation}
is the Fermi velocity.    
Consequently, the integration in Eq.~(\ref{orig}) over the wave vector
can be also performed relative to the each Dirac cone vertex,   
\begin{equation}
g_{0}(E)\approx %
\frac{2}{S_{BZ}}\int\frac{E}{E^{2}-E^{2}(\bm{k})}d\bm{k},%
\qquad E\ll 3t,
\label{mod}
\end{equation}    
where the factor of $2$ reflects the existence of two Dirac cones in the
spectrum. However, due to the mutual overlap between respective cones,
the integration in (\ref{mod}) can not be done over the entire
Brillouin zone within the linear approximation for the dispersion
relation. The corresponding cutoff magnitude of the wave vector is
determined by the sum rule, 
\begin{equation}
\frac{4\pi}{S_{BZ}}\int_{0}^{k_{max}}dk=1,
\end{equation}
which yields
\begin{equation}
k_{max}=\frac{2\sqrt{\pi}}{\sqrt{\sqrt{3}}a}.
\end{equation}
Then, the integration can be performed exactly,
\begin{multline}
g_{0}(E)\approx \frac{4\pi}{S_{BZ}}\int_{0}^{k_{max}}%
\frac{E}{E^{2}-(v_{F}k)^{2}}kdk=\int_{0}^{1}%
\frac{E}{E^{2}-\sqrt{3}\pi t^{2}x}dx= \\ =\frac{\epsilon}{W}%
\left[\ln\left(\frac{\epsilon^{2}}{1-\epsilon^{2}}\right)%
-i\pi\sgn \epsilon\right],%
\qquad E<W,
\end{multline}
where
\begin{equation}
W=\sqrt{\pi\sqrt{3}}t, \label{bw}
\end{equation}
is the bandwidth parameter for the pure Dirac spectrum.

\end{document}